\documentclass[preprint,floatfix,nofootinbib,superscriptaddress]{elsarticle}

\usepackage{graphicx,subfig,epsfig,epsf,color}
\usepackage{amssymb,amsmath}
\usepackage{slashed}
\usepackage{feynmf}
\usepackage{array}
\usepackage{hyperref}

\newcolumntype{C}[1]{>{\centering\let\newline\\\arraybackslash\hspace{0pt}}m{#1}}

\newcommand{\beq}{\begin{equation}}
\newcommand{\eeq}{\end{equation}}
\newcommand{\bea}{\begin{eqnarray}}
\newcommand{\eea}{\end{eqnarray}}
\newcommand{\beas}{\begin{eqnarray*}}
\newcommand{\eeas}{\end{eqnarray*}}
\newcommand{\bcr}{\begin{center}}

\def\Re{{\cal R \mskip-4mu \lower.1ex \hbox{\it e}\,}}
\def\Im{{\cal I \mskip-5mu \lower.1ex \hbox{\it m}\,}}

\def\tev{\,{\ifmmode\mathrm {TeV}\else TeV\fi}}
\def\gev{\,{\ifmmode\mathrm {GeV}\else GeV\fi}}
\def\mev{\,{\ifmmode\mathrm {MeV}\else MeV\fi}}
\def\to{\rightarrow}

\begin{document}

\def\issue(#1,#2,#3){#1 (#3) #2} 
\def\APP(#1,#2,#3){Acta Phys.\ Polon.\ \issue(#1,#2,#3)}
\def\ARNPS(#1,#2,#3){Ann.\ Rev.\ Nucl.\ Part.\ Sci.\ \issue(#1,#2,#3)}
\def\CPC(#1,#2,#3){comp.\ Phys.\ comm.\ \issue(#1,#2,#3)}
\def\CIP(#1,#2,#3){comput.\ Phys.\ \issue(#1,#2,#3)}
\def\EPJC(#1,#2,#3){Eur.\ Phys.\ J.\ C\ \issue(#1,#2,#3)}
\def\EPJD(#1,#2,#3){Eur.\ Phys.\ J. Direct\ C\ \issue(#1,#2,#3)}
\def\IEEETNS(#1,#2,#3){IEEE Trans.\ Nucl.\ Sci.\ \issue(#1,#2,#3)}
\def\IJMP(#1,#2,#3){Int.\ J.\ Mod.\ Phys. \issue(#1,#2,#3)}
\def\JHEP(#1,#2,#3){J.\ High Energy Physics \issue(#1,#2,#3)}
\def\JPG(#1,#2,#3){J.\ Phys.\ G \issue(#1,#2,#3)}
\def\MPL(#1,#2,#3){Mod.\ Phys.\ Lett.\ \issue(#1,#2,#3)}
\def\NP(#1,#2,#3){Nucl.\ Phys.\ \issue(#1,#2,#3)}
\def\NIM(#1,#2,#3){Nucl.\ Instrum.\ Meth.\ \issue(#1,#2,#3)}
\def\PL(#1,#2,#3){Phys.\ Lett.\ \issue(#1,#2,#3)}
\def\PRD(#1,#2,#3){Phys.\ Rev.\ D \issue(#1,#2,#3)}
\def\PRL(#1,#2,#3){Phys.\ Rev.\ Lett.\ \issue(#1,#2,#3)}
\def\PTP(#1,#2,#3){Progs.\ Theo.\ Phys. \ \issue(#1,#2,#3)}
\def\RMP(#1,#2,#3){Rev.\ Mod.\ Phys.\ \issue(#1,#2,#3)}
\def\SJNP(#1,#2,#3){Sov.\ J. Nucl.\ Phys.\ \issue(#1,#2,#3)}

\bibliographystyle{elsarticle}

\title{$q$-deformed Einstein's Model to Describe Specific Heat of Solid}

\author[bits]{Atanu~Guha\corref{cor1}}
\ead{am.atanu@gmail.com/ p20140401@goa.bits-pilani.ac.in}

\author[bits]{Prasanta~Kumar~Das\corref{cor1}}
\ead{pdas@goa.bits-pilani.ac.in}

\cortext[cor1]{Corresponding author}
\cortext[cor2]{Co-author}

\address[bits]{Birla Institute of Technology and Science-Pilani, Department of Physics, 
Goa campus, NH-17B, Zuarinagar, Goa-403726, India}

\date{\today}

\begin{abstract} 
Realistic phenomena can be described more appropriately using generalized canonical ensemble,  with proper parameter sets involved. 
 We have generalized the Einstein's theory for specific 
 heat of solid in Tsallis statistics, where the temperature fluctuation 
 is introduced into the theory via the fluctuation parameter $q$. At low temperature the Einstein's curve of the
 specific heat in the nonextensive Tsallis scenario exactly lies on the experimental data points. 
 Consequently this $q$-modified Einstein's curve is found to be overlapping with the one predicted by Debye.
 Considering only the temperature fluctuation effect(even without considering more than one
 mode of vibration is being triggered) we found that the $C_V$ vs $T$ curve is as 
 good as obtained by considering the different modes of vibration as suggested by Debye. Generalizing
 the Einstein's theory in Tsallis statistics we found that a unique value 
 of the Einstein temperature $\theta_E$ along with a temperature dependent deformation
 parameter $q(T)$, can well describe the phenomena of specific heat of solid 
 i.e. the theory is equivalent to Debye's theory with a temperature dependent $\theta_D$.

\end{abstract}

\maketitle

\section{Introduction}

 Statistical mechanics has been proved to be one of the most powerful tools in various 
 domains over the last century. It has been successfully used not only in 
 different branch of physics (e.g. condensed matter physics, high energy physics, Astrophysics
 etc.), but in different areas beyond those (e.g. share price dynamics, dynamics related to traffic 
 control etc). The results predicted by the Statistical Mechanics have been found to be in 
 good agreement with the experiments. The important fact here is that one can predict the macroscopic properties of the system without having much detailed knowledge of each and 
 every microstate of the system, but based on the notion of the statistical average of 
 the microscopic properties. 
 
  Attempts have been made to generalize this important tool (i.e. statistical mechanics) in recent years\cite{Tsallis,Sumiyoshi, ugur}. This generalized statistical techniques(popularly known as superstatistics or nonextensive statistics (in Tsallis statistics $q$ being the deformation parameter)) have been applied to a wide range of complex systems, e.g., hydrodynamic turbulence, defect turbulence, share price dynamics, random matrix theory, random networks, wind velocity fluctuations, hydroclimatic fluctuations, the statistics of train departure delays and models of the metastatic cascade in cancerous systems \cite{Tsallis5, Tsallis6, Plastino, Plastino2, Plastino3, Fevzi2, Fevzi3}. 
 
 In this approach, the key parameter is the inverse temperature parameter $\beta~(=1/k_B T)$ 
 which exhibits fluctuations(e.g. in turbulent system with energy dissipation) on a large time scale. These type of complex systems can be modelled by superposition of ordinary statistical mechanics with varying temperature 
 parameters, which is called superstatistics or deformed statistics to sum up. The stationary distributions
 of deformed/superstatistical  systems differs from the usual Boltzmann-type statistical mechanics 
 and they can disclose themselves into asymptotic power laws or some different functional forms in the energy $E$ \cite{cbeck1, cbeck2, cbeck3}.

 By using the non-extensive statistical methods one can incorporate the fact of temperature 
 fluctuations and the proceeding sections are devoted mainly for an attempt to give more 
 insight on it \cite{ugur, Tsallis2, cohen, Tsallis3}. This approach deals with the fluctuation 
 parameter $q$ which corresponds to the degree of the temperature fluctuation effect to 
 the concerned system. Detailed theoretical studies for nonextensive systems can be found in \cite{rajagopal, rajagopal2, rajagopal3}. In this formalism we can treat our normal Boltzmann-Gibbs statistics 
 as a special case of this generalized one, where temperature fluctuation effects are 
 negligible, corresponds to $q=1.0$. More deviation of $q$ from the value $1.0$ denotes 
 a system with more fluctuating temperature. Various works related to this generalized or nonextensive 
 statistics have been reported in different phenomena \cite{ugur, cbeck5, Tsallis4, Tsallis10, Atanu}. More works have been done on theoretical understanding as well as different applications \cite{martinez, Tsallis20, Tsallis21, Ramshaw, Tsallis22, Tsallis23, rajagopal4, rajagopal5, abe1, abe2, Tsallis24, Tsallis25, plastino10, hamity, chimento, boghosian, mendes10, abe4, mendes11, plastino11, abe5}. 
 
\section{Temperature fluctuation and the modified entropy}

 The phenomena of temperature fluctuation can be interpreted physically as the deformation 
 of a ideal canonical ensemble to a more realistic case. Ideal canonical ensemble is supposed 
 to be the statistical ensemble that represents the possible states of a mechanical system 
 in thermal equilibrium with a heat bath at a fixed temperature, say $T$. Consequently 
 each and every cell(very small identical portions of the system) will be at temperature $T$.
 To make it more realistic we can think about a modified canonical system which is in 
 thermal equilibrium with a heat bath at a fixed temperature $T$ but there will be a small 
 variation in temperature in different cells, say between $T-\delta T$ to $T+\delta T$, 
 though the average temperature of the system will be $T$ still.
 
 A connection between the entropy ($s$) and the number of microstates ($\Omega$) of a system
 can be derived intuitively as follows. We know the entropy is a measure of the degree of randomness of 
 a system i.e. the number of possible microscopic configuration. The only thing we 
 can infer clearly is that, they both $s$ and $\Omega$ will increase (or decrease) together. 
 Assuming $s=f(\Omega)$ and noting that the entropy is additive and the 
 number of microstates is multiplicative, a simple calculation yields  $ s=k_B \ln \Omega $. 
 
 A more general connection between $s$ and $\Omega$ can be made (in the context of 
 generalized/deformed statistics) which will deform the fundamental relation as 
\bea
s=f(\Omega^q),
\eea
where the deformation parameter $q>0$. Subsequently, the generalized entropy can be shown to take the 
following form. 
\bea
s_q=k_B \ln_q \Omega
\eea
where the generalized log function($\ln_q \Omega$) is defined as
\bea
\ln_q \Omega = \frac{\Omega^{1-q}-1}{1-q}.
\label{eqn:lnq}
\eea
Consequently the generalized exponential function becomes
\bea
e_q^x=\left[1+(1-q) x \right]^\frac{1}{1-q}.
\label{eqn:expq}
\eea

 Therefore $q$-modified Shannon entropy takes the following form
\bea
s_q= k_B \frac{\sum p_i^q -1}{1-q}
\eea

 Extremizing $s_q$ subject to suitable constraints yields more general canonical ensembles(see  \ref{entropy_optimization}),
where the probability to observe a microstate with energy $\epsilon_i$ is given by: \cite{Tsallis, Tsallis12, oikonomou}  
\bea
p_i = \frac{e_q^{- \beta' \epsilon_i}}{Z_q} = \frac{1}{Z_q} \left[ 1 - (1-q) \beta' \epsilon_i \right]^{\frac{1}{1-q}}
\eea
with partition function $Z_q$ and inverse temperature parameter $\beta=\frac{1}{k_B T}$. Also $\beta'$ is the $q$-modified quantity and is given by \cite{Tsallis, Tsallis12} 
\bea
\beta'=\frac{\beta}{\sum_i p_i^q+(1-q) \beta u_q}=\frac{\beta}{Z_q^{1-q}+(1-q) \beta u_q}
\eea 

 with $q$-generalized average energy 
\bea
u_q=\frac{\sum_i \epsilon_i p_i^q}{\sum_i p_i^q}
\eea 

The $q$-deformed/generalized exponential function can be expanded as follows 
\bea
e_q^x &=& \left[1+(1-q) x \right]^{\frac{1}{1-q}} \nonumber \\
&=& 1+x+ q \frac{x^2}{2!}+ q (2q-1) \frac{x^3}{3!}+ q (2q-1) (3q-2) \frac{x^4}{4!}+\cdots
\eea
where $x=- \beta \epsilon_i= - \frac{\epsilon_i}{k_B T}$. The $q$ factors in the expansion, 
which can be absorbed in $T$, will account for the temperature fluctuation of the system.  
By setting $q=1.0$ (which corresponds to zero or negligible temperature fluctuation) one gets
back the normal Boltzmann-Gibbs statistics.
 


\section{Specific heat of solid in the light of Tsallis statistics }

 From the previous discussion it is evident that we can use this generalized/deformed statistical 
 mechanics wherever the system is subjected to some kind of temperature fluctuation. If the 
 temperature fluctuation effect is not negligible enough to disclose itself, then definitely
 there will be some deviation from the ideal phenomena.

\subsection{Einstein's theory of specific heat}

 Einstein viewed the specific heat of solid as an effect of the vibrations of the solid. He treated the 
 atoms in a $N$-atoms solid (e.g. crystal) as $N$ 3-D simple harmonic oscillators, each of which is 
 vibrating with the common frequency $\nu_E$. The magnitude of $\nu_E$ depends on the strength 
 of the restoring force acting on each atom(which in this case considered to be the same 
 for each atom). 
 Now we know a solid of $N$ atoms is equivalent to $3\rm{N}$ 1-D harmonic oscillators. 
 So we can treat each atom as a collection of 3 vibrating 1-D harmonic oscillators and all 
 the $3\rm{N}$ 1-D oscillators are vibrating with a common frequency $\nu_E$. 
 This kind of treatment is a gross approximation as the lattice vibrations, in reality, are very complicated coupled 
 oscillations \cite{Einstein, Reif, Patharia, Kittel}. 

 The energy levels of the one-dimensional harmonic oscillator can be written as 
\bea 
\epsilon_n=  \left(n+\frac{1}{2}\right) h \nu_E
\eea 
where $h$ is the Planck constant and $n = 0, 1, 2, \cdots$.  In the treatment of canonical 
ensemble all such 1-D oscillators are in thermal equilibrium (say, at temperature $T$). 
For a single 1-D oscillator, the partition function($Z$) can be written as  
\bea 
Z &=& \sum_{n=0}^{\infty} \exp \left(-\beta \epsilon_n \right) \nonumber \\
&=& \sum_{n=0}^{\infty} \exp \left\lbrace-\beta \left(n+\frac{1}{2}\right) h \nu_E \right\rbrace \nonumber \\ 
&=& e^{-x/2} \sum_{n=0}^{\infty} e^{-n x} = \frac{e^{-x/2}}{1-e^{-x}}, ~~\rm{where}~ x= \beta h \nu_E
\eea
In above, we have used the fact that $\sum_{n=0}^{\infty} x^n = \frac{1}{1-x}$. Accordingly, 
the mean energy of a single oscillator is found to be 
\bea 
 u &=& \sum_{n=0}^{\infty} \frac{\epsilon_n \exp \left(-\beta \epsilon_n \right)}{Z} \nonumber \\
 &=& -\frac{\partial \left( \ln Z \right)}{\partial \beta} = \frac{\partial}{\partial \beta} \left\lbrace \frac{\beta h \nu_E}{2} + \ln \left(1- e^{-\beta h \nu_E} \right) \right\rbrace = \frac{h \nu_E}{2} + \frac{h \nu_E}{e^{\beta h \nu_E}-1}
\eea
Here $\frac{h \nu_E}{2}$ is the zero point energy. The energy of the $3N$ 1-D oscillators 
in the $N$-atom solid is given by
\bea
 U = 3N u = 3N \left( \frac{h \nu_E}{2} + \frac{h \nu_E}{e^{\beta h \nu_E}-1} \right)
\eea

We write the heat capacity at constant volume as
\bea 
 C_V &=& \left(\frac{\partial U}{\partial T} \right)_V \nonumber \\
&=& 3N \left(\frac{\partial U}{\partial \beta} \right)_V \left(\frac{\partial \beta}{\partial T} \right) 
= 3 N k_B \frac{x^2 e^x}{\left(e^x -1 \right)^2}, ~~\rm{with}~ x =\frac{h \nu_E}{k_B T}
= \frac{\theta_E}{T}.
\label{eq:cveinstein}
\eea
In above $\theta_E$ is called the "Einstein temperature". It is different for different 
solid and reflects the lattice rigidity. Now when the temperature is very high i.e. 
$T \gg \theta_E$ (i.e. $x \ll 1$), the Einstein heat capacity reduces to $C_V = 3N k_B$ 
which is the Dulong and Petit law. We set $e^x \sim 1+x$ in the denominator of the specific heat 
expression above (Eq. $\ref{eq:cveinstein}$)for small $x$ while getting the Dulong and Petit law. 
 
When the temperature  is low i.e. when $T \ll \theta_E$ (i.e., $x \gg 1$),  the Einstein 
specific heat $C_V \rightarrow  0$ as $T \rightarrow 0$.  This is obtained by setting 
$ e^x-1 \sim e^x $ in the denominator of the specific heat expression 
(Eq. $\ref{eq:cveinstein}$) for large $x$. 
This is also  a requirement which follows from the third law of thermodynamics.

\subsection{Debye's modification to Einstein's model of specific heat}
 Debye did a major improvement of the Einstein's model. He treated the coupled vibrations of
 the solid in terms of $3N$ normal modes of vibration of the entire solid, each with its own frequency. So in his treatment the lattice vibrations are equivalent to $3N$ independent harmonic oscillators 
 with different normal mode frequencies. The crystal for low frequency vibrations can be treated as a homogeneous elastic medium. For the low frequencies, the wavelength $\lambda \gg a $, 
 where $a$ is the atomic spacing. The normal modes mentioned above, are defined as the 
frequencies of the standing waves. The number of normal modes with the frequency ranges between $\nu$ and $\nu+d \nu$ in such a
medium \cite{Debye, Reif, Patharia, Kittel}.
\bea
 g(\nu) d \nu = \frac{4 \pi V \nu^2}{v^3} d\nu 
\label{normal_mode} 
\eea
where $V$ is the volume of the crystal and $v$ is the propagation velocity of the wave in 
solid. The expression mentioned above(Eq.(\ref{normal_mode})), is applicable only to the low frequency vibrations in a crystal. The approximation used by Debye to make the form useful is that the expression(Eq.(\ref{normal_mode})) applies to all frequencies, and he established the concept of a maximum 
frequency $\nu_D$ (the Debye frequency) to ensure the total number of modes to be $3N$, i.e., 
$ \int_0^{\nu_D} g(\nu) d \nu = 3 N $. 
Now we can integrate over all of the frequencies to find the internal energy of the 
crystal with the approximation discussed above, as suggested by Debye. Hence in Debye's theory the heat capacity is given by 
\bea
 C_V = \left(\frac{\partial U}{\partial T}\right)_V = 3N k_B \left\lbrace \frac{3}{x_D^3} \int_0^{x_D} \frac{x^4 e^x }{\left(e^x -1 \right)^2} dx \right\rbrace
 \label{eq:cvdebye}
\eea
where $x =\frac{h \nu}{k_B T}$, and $x_D =\frac{h \nu_D}{k_B T} = \frac{\theta_D}{T}$, where $\theta_D$
is the Debye temperature. Clearly $C_V$ depends on $\theta_D$. Analytically we cannot evaluate the integral, it has to be done numerically. 
At high temperatures ($T \gg \theta_D,~ x_D \ll 1$), we can give a compact form to the integrand, rewriting in the following way: 
\bea
\frac{x^4 e^x}{\left(e^x - 1\right)^2} 
= \frac{x^4}{\left(e^x - 1 \right) \left(1- e^{-x} \right)}
 =\frac{x^4}{2 \left(\cosh(x) - 1 \right)} = \frac{x^4}{2 \left( \frac{x^2}{2!} + \frac{x^4}{4!} + \cdots \right)}
\eea

 Keeping only the $x^2$ term in the denominator we get
 \bea 
 C_V = 3N k_B \left\lbrace \frac{3}{x_D^3} \int_0^{x_D} x^2 dx \right\rbrace = 3N k_B
\eea 
which is the Dulong and Petit law. To determine the heat capacity at the low temperature 
limit ($T \ll \theta_D,~ x_D \gg 1$), we see that the integrand (in Eq.(\ref{eq:cvdebye})) 
tends towards zero rapidly for large $x$. We replace the upper limit by $\infty$ and turn 
the integral into a standard integral to give
\bea 
 C_V = 3N k_B \left( \frac{T}{\theta_D} \right)^3 \left\lbrace 3 \int_0^{\infty} \frac{x^4 e^x}{\left(e^x - 1\right)^2} \right\rbrace 
 = \frac{12}{5} \pi^4 N k_B \left(\frac{T}{\theta_D}\right)^3
\eea 
Thus we see that Debye heat capacity varies as $T^3$ at low temperatures, in agreement 
with experimental observation, which is a remarkable improvement of Einstein’s theory. 
 
\subsection{Modification of Einstein's theory of specific heat using Tsallis statistics} 	

 Here, with the understanding of the nonextensive Tsallis statistics as the key element which takes care for the temperature fluctuation caused by the nearest neighbour interaction of the atoms, we would like to propose the following. Let us view the $N$-atom solid as a collection of $N$ 
$3$-dimensional(3D) $q$-deformed harmonic oscillators, 
which are equivalent to $3N$ $1$-dimensional(1D) $q$-deformed harmonic oscillators, each is vibrating with 
angular frequency $\omega$. The average energy of each of these $q$-deformed oscillators 
is given by \cite{Tsallis, Sumiyoshi, Tsallis12, Swamy} 
\bea
<\epsilon>_q = \frac{\sum_i \epsilon_i~ p_i^q}{\sum_i p_i^q}= u_q
\eea
where the probability $p_i$ for a particular energy eigenstate $\epsilon_i$ is given by
\bea
p_i= \frac{\left[1-\left(1-q\right)\beta' \epsilon_i\right]^{\frac{1}{1-q}}}{Z_q}=\frac{e_q^{-\beta' \epsilon_i}}{Z_q}
\eea
and the $q$-deformed partition function $Z_q=\sum_i e_q^{-\beta' \epsilon_i}$

 with 
 \bea
\beta'=\frac{\beta}{\sum_i p_i^q+(1-q) \beta u_q}=\frac{\beta}{Z_q^{1-q}+(1-q) \beta u_q}
\eea

 Now the total internal energy of the system becomes \cite{Tsallis, Swamy, guo}
\bea
U_q=-\frac{\partial}{\partial \beta} \ln_q Z_q =  3N <\epsilon>_q
\eea 
The molar specific heat capacity of the system at constant volume
\bea
C_V=\frac{\partial U_q}{\partial T}=3N \frac{\partial <\epsilon>_q}{\partial T}
\label{cv1}
\eea
Now for the $q$-deformed harmonic oscillators the energy eigenvalues are
\bea
\epsilon_n=\left(n+\frac{1}{2}\right) \hbar \omega= k_B T \left(n+\frac{1}{2}\right) x
\eea
where $x$ is new variable defined as $\frac{\hbar \omega}{k_B T}$ for convenience and consequently we got the following relation
\bea
x \frac{\partial}{\partial x}= -T \frac{\partial}{\partial T}
\eea
Substituting all these in Eq.[\ref{cv1}] we get the following 
\bea
C_V=3N \frac{\partial}{\partial T} \left[k_B T \cdot \frac{\sum_n \left(n+\frac{1}{2}\right)x \left\lbrace e_q^{-\frac{\beta'}{\beta} \left(n+\frac{1}{2}\right)x}\right\rbrace^q}{\sum_n \left\lbrace e_q^{-\frac{\beta'}{\beta}  \left(n+\frac{1}{2}\right)x}\right\rbrace^q} \right]
\eea
Further reduction gives the following simplified form
\bea
\frac{C_V}{3R} = \frac{\sum_n \left(n+\frac{1}{2}\right)x \left\lbrace e_q^{-\frac{\beta'}{\beta} \left(n+\frac{1}{2}\right)x}\right\rbrace^q}{\sum_n \left\lbrace e_q^{-\frac{\beta'}{\beta}  \left(n+\frac{1}{2}\right)x}\right\rbrace^q} - x \frac{\partial}{\partial x} \left[\frac{\sum_n \left(n+\frac{1}{2}\right)x \left\lbrace e_q^{- \frac{\beta'}{\beta}  \left(n+\frac{1}{2}\right)x}\right\rbrace^q}{\sum_n \left\lbrace e_q^{-\frac{\beta'}{\beta}  \left(n+\frac{1}{2}\right)x}\right\rbrace^q} \right]
\label{cvfinal}
\eea
where $R=N \cdot k_B$ stands for the universal gas constant.
 
 For $q=1$(i.e., undeformed scenario) Eq.(\ref{cvfinal}) becomes Eq.(\ref{eq:cveinstein}), the Einstein's expression of specific heat.

\protect\label{q_modified_einstein}

\section{Numerical Analysis}
 In the  following discussions Subsec. \ref{numerical_einstein} and \ref{numerical_debye}, we explore 
 several interesting features of $q$-deformed Einstein's  models of specific heat.

\subsection{Aspects of $q$-deformed Einstein's model and its phenomenology} 

\protect\label{numerical_einstein}

\begin{figure}[htbp]
\includegraphics[width=10cm]{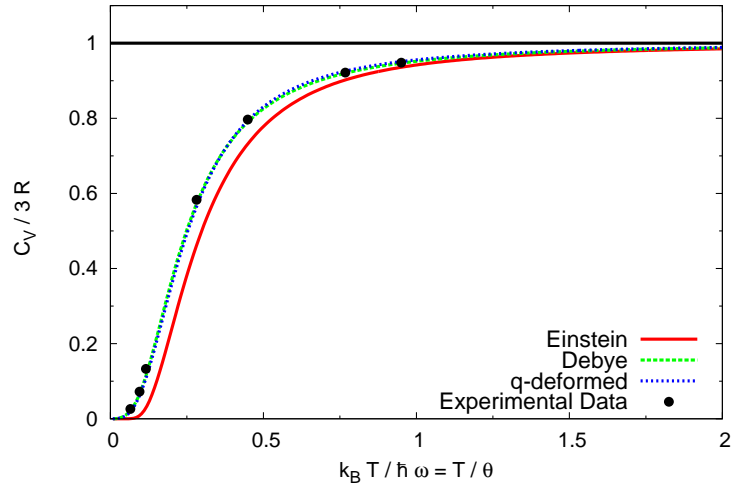}
\caption{Variation of $C_V$ with temperature in the Einstein's model
(with $\theta=\theta_E$), Debye's modification of Einstein model(with $\theta=\theta_D$) 
and $q$-deformed Einstein's model($\theta=\theta_E$) is shown. For Copper, 
experimental points lies on the curve predicted by Debye \cite{white, bloom, stewart, Patharia}
and as well as on the curve predicted by $q$-deformed Einstein's oscillator model. 
The horizontal curve on the top  denotes the Dulong$-$Petit law($C_V/3 R  \sim 1$) at high 
temperature.}
\label{cve}
\end{figure}

 In Fig.[\ref{cve}] we have shown $\frac{C_V}{3R}$ as a function of 
$\frac{1}{x}(=\frac{T}{\theta})$ using Einstein's theory, 
Debye's modification and $q$-deformed Einstein's theory. 
The following observations are in order.
\begin{itemize}
\item The horizontal curve at $\frac{C_V}{3R} = 1$ corresponds to Dulong$-$Petit's law of specific 
heat at sufficiently high temperature.
\item The small filled circles correspond to experimental data points of the specific heat of Copper($Cu$) \cite{white, bloom, stewart, Patharia}.
\item The blue dotted curve, which matches quite well with the experimental data point and Debye's curve, 
corresponds to the specific heat variation in $q$ deformed scenario using Eq.(\ref{cvfinal}). Instead of using a constant $q$-value over the wide range of temperature, 
we have used the temperature ($T$) dependent deformation parameter $q$, 
(a well known exponential decay function with $3$ free parameter, namely $q_0, A ~\rm{and}~ t $)) 
given by (Eq.(\ref{qtapp}))
\bea
q(T)=q_0 + A \exp {(- \frac{1}{t}\frac{k_B T}{\hbar \omega})}
\label{fit_q}
\eea 
 with $q_0=1.0, A=0.08 ~\rm{and}~ t=0.2 $ (See \ref{fit_Einstein} for the related discussion).
Here $q=q_0=1.0$ the undeformed case; $A$ and $t$ are the parameters 
 which can be determined for a material by fitting experimental data. Also we used numerical derivative of $q(T)$ in Eq.(\ref{cvfinal}) for convenience.
 
 \item As $T \to \theta_E$, $q(T) \to q_0 (=1)$ i.e. at temperature higher or above 
 $\theta_E$, the fluctuation in the deformation parameter $\delta q(= q(T) - q_0) \to 0$ 
 can be neglected in this phenomenon of specific heat of solids.
 \item At other temperatures(moderate, low and very low temperatures), the curve due to $q$-deformed Einstein's oscillators  
 is in perfect agreement with the Debye curve and the experimental data. Note that here we considered only 
 one single excitation mode of the oscillators (as Einstein did) and treat each atom as the 
 $q$-deformed oscillator. This mode represents the fundamental frequency of the oscillators
 for a specific material. This is completely different from the Debye's approach, in which 
 the specific heat is viewed as the vibration of multiple excitation modes of the oscillators.
 \end{itemize}

 \begin{itemize}
\item We see that by considering only the temperature fluctuation effect(even without considering more than one mode of vibration
is being triggered) we can achieve the desired result (i.e. the specific heat prediction 
in the  $q$-deformed Einstein's scenario matches exactly with the experimental data points) which is as good as obtained by 
considering the different modes of vibration as suggested by Debye. 

\item By fitting the experimental data for Copper(Cu) with Einstein's theory, Debye's theory and $q$-deformed Einstein's theory in Fig.[\ref{cve}] we also found that, $ \theta_E \sim 0.77 \theta_D \approx 240~\rm{K} $ in room temperature, as $\theta_D =310~\rm{K}$ for Copper in room temperature \cite{powell}. This is consistent with the theoretical expectation $ \theta_E \sim 0.8 \theta_D$ \cite{james, daniel, Terrell}.
\end{itemize}
 
\subsection{Analysis on $q$-deformed Einstein's theory Vs Debye's theory}

\protect\label{numerical_debye}

 Experimental observation shows that the  Debye temperature($\theta_D$) (which is 
 nothing but the rescaling of the Debye frequency($\nu_D$) for a specific material) depends 
 on the temperature of the material\cite{white,bloom,flubacher,shulman}. Experimental data 
 of the heat capacity for a specific material at very low temperature matches with the curve 
 predicted by Debye with a higher $\theta_D$, whereas for the same material at room 
 temperature, the matching requires lower value of $\theta_D$. 
\begin{figure}[!ht]
  \centering
  \subfloat[$T= 0\rm{-}400 \rm{K}$]{\includegraphics[width=0.49\textwidth]{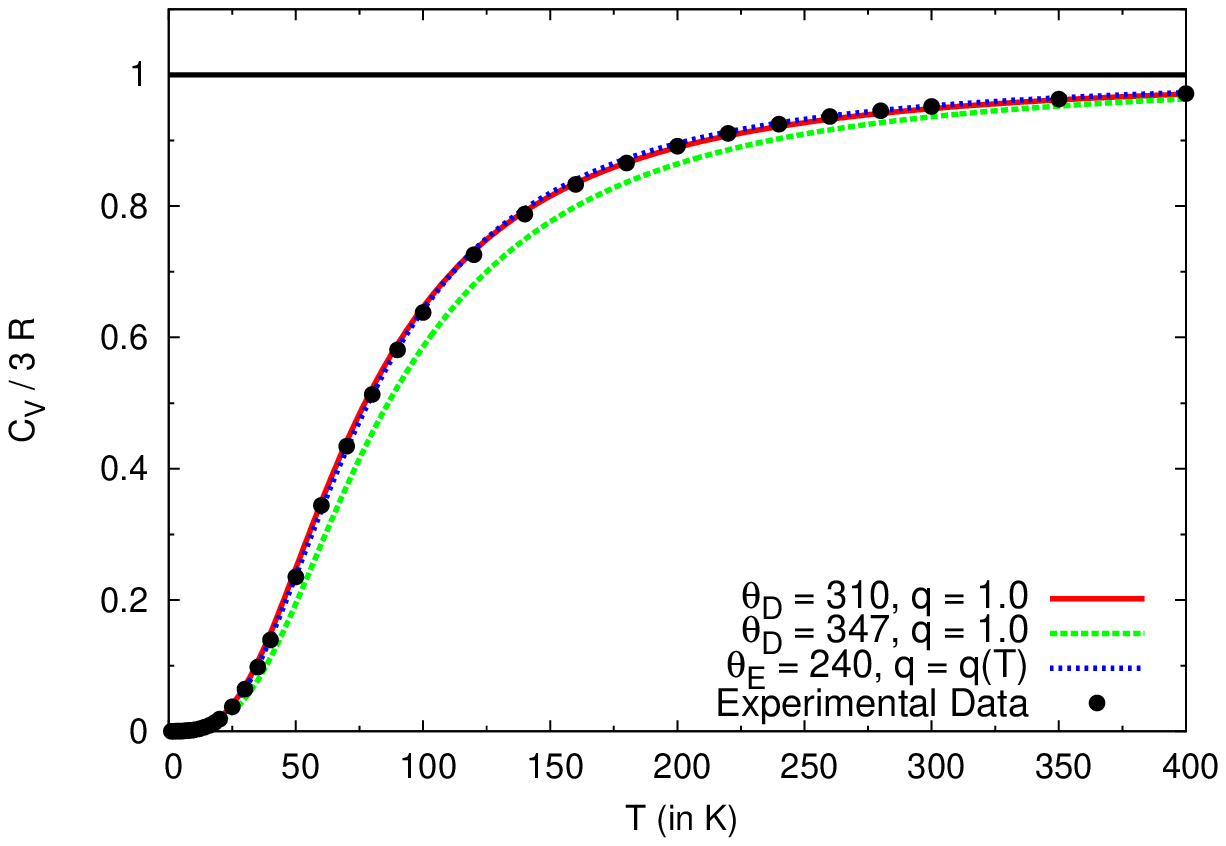}}
  \hfill
  \subfloat[$T= 10\rm{-}70 \rm{K}$]{\includegraphics[width=0.49\textwidth]{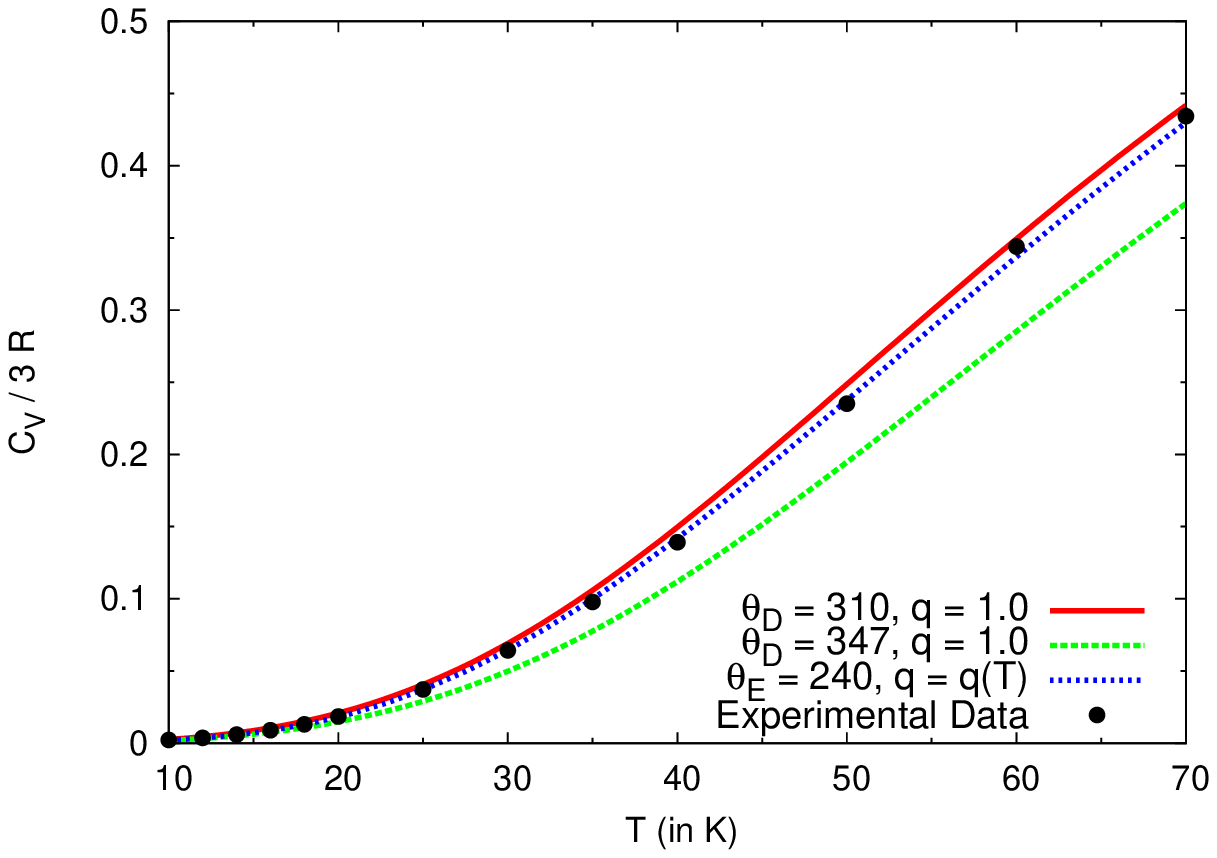}}
  \\
  \centering
  \subfloat[$T= 0\rm{-}30 \rm{K}$]{\includegraphics[width=0.49\textwidth]{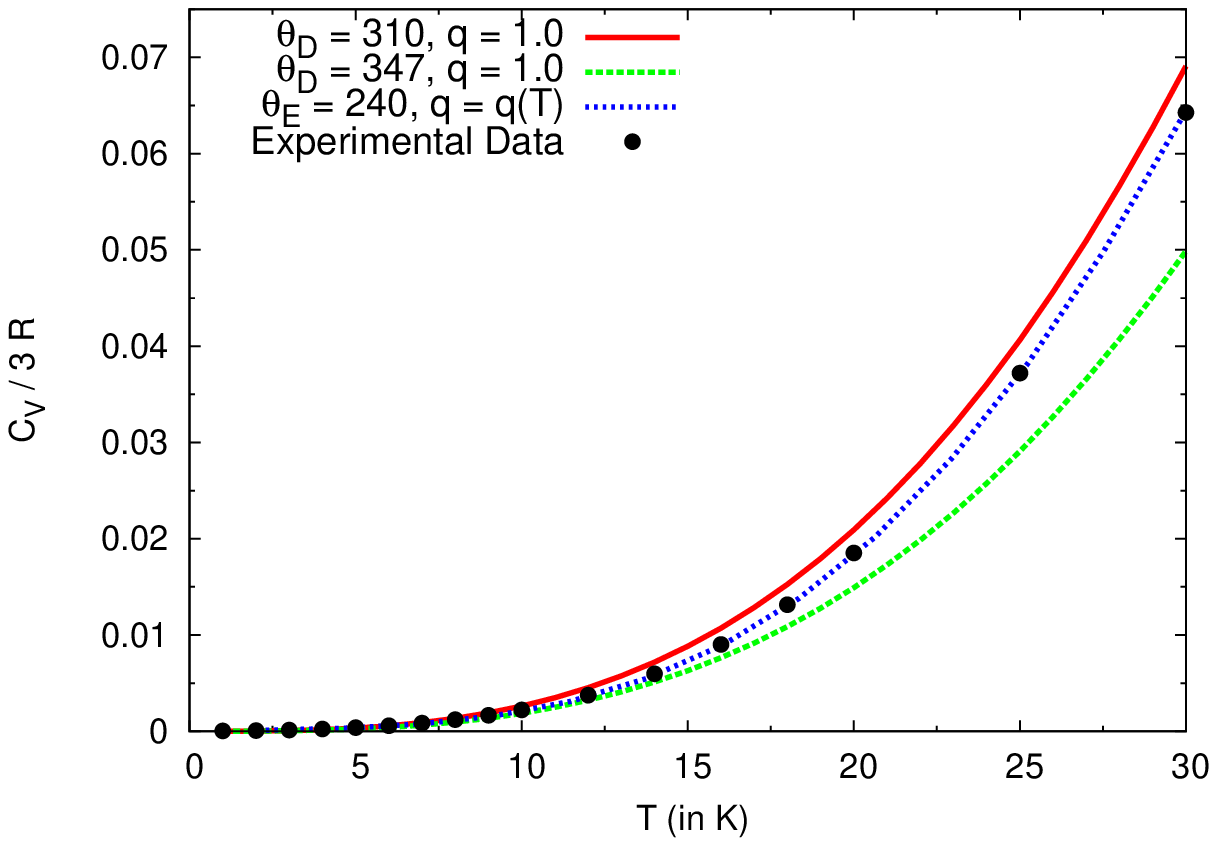}}
  \hfill
  \subfloat[$T= 0\rm{-}15 \rm{K}$]{\includegraphics[width=0.49\textwidth]{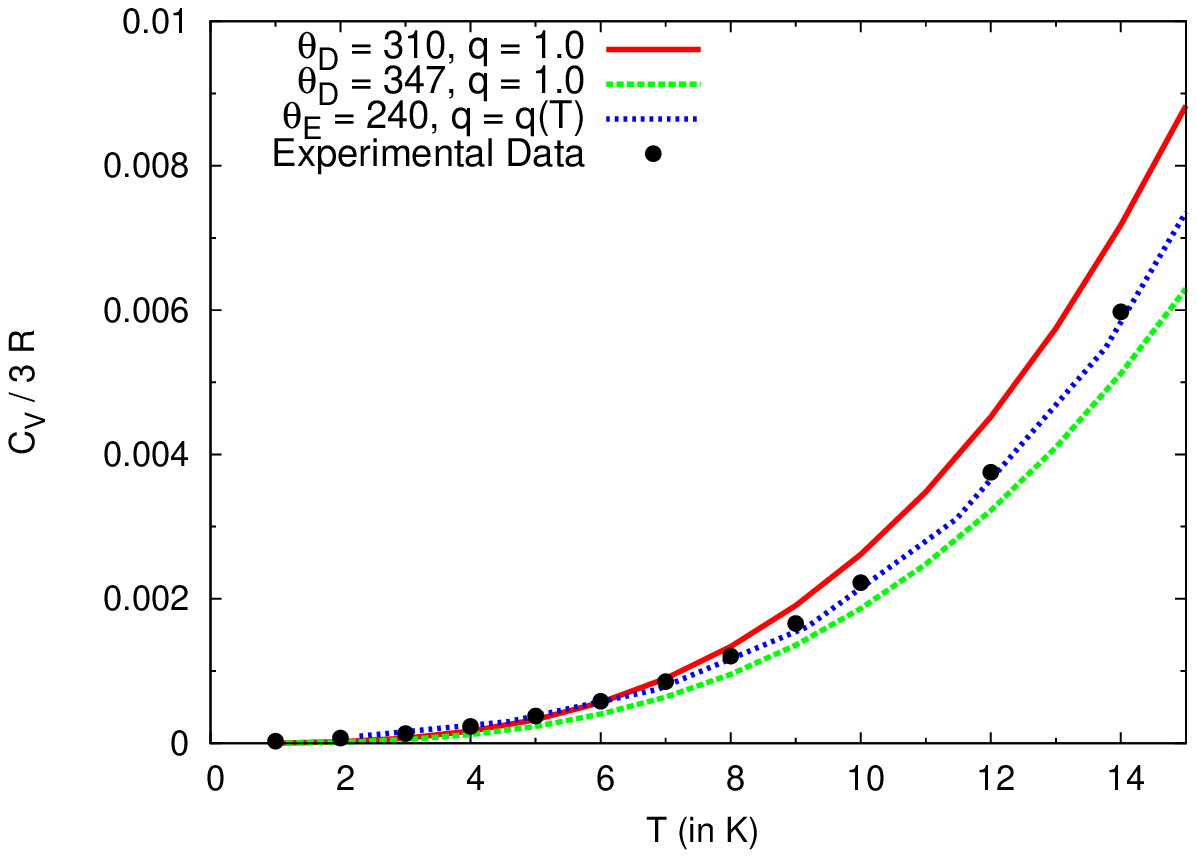}}
  \caption{\it Variation of $C_V/3R$ (for $Cu$) with temperature in the Debye's model and 
$q$-deformed Einstein's model at very low temperature($0-100 ~\rm{K}$) is shown along with 
the experimental data points \cite{white, stewart}. The horizontal line corresponds to 
$C_V \sim 3 R$.}

\label{qdebye}
\end{figure}
 For example, at low 
 temperature ($\sim 0 ~\rm{K}$) the values of $\theta_D$ for Copper(Cu) and Aluminium(Al) 
 are $347 ~\rm{K}$ and $433 ~\rm{K}$ respectively \cite{stewart, tari}; whereas, at 
 room temperature($\sim 300 ~\rm{K}$) the values of $\theta_D$ for Copper(Cu) and 
 Aluminium(Al) are $310 ~\rm{K}$ and $390 ~\rm{K}$ respectively \cite{powell}. Hence, the Debye's temperature $\theta_D$ is a temperature dependent quantity i.e., $\theta_D = \theta_D(T)$.

  We propose an alternate scenario(i.e., the $q$-deformed Einstein's model, mentioned above), in which $\theta_E$ is assumed to be 
 independent of the temperature. 
 We consider the same form of $q(T)$  
 \bea
 q(T)=q_0 + A \exp {(- \frac{1}{t}\frac{k_B T}{\hbar \omega})}
\label{q_fit_debye} 
\eea
which at high temperature (i.e. $T \to \infty$) goes to $1$.

 In Fig.[\ref{qdebye}] we have plotted 
$\frac{C_V}{3R}$ as a function of Temperature(in K) for Copper(Cu) with $\theta_D=310 ~\rm{K} ~\rm{and}~347 ~\rm{K}$ in Debye's 
theory (using Eq.(\ref{eq:cvdebye})) and for the $q$-deformed Einstein's theory 
(using Eq.(\ref{cvfinal})) with $\theta_E=240 ~\rm{K}$ as well. With $q=q(T)$ 
 Eq.(\ref{fit_q}) and (\ref{q_fit_debye}) there is a smooth transition between the two curves and one requires 
 the following setting of the three free parameters i.e. $q_0=1.0, A=0.08 ~\rm{and}~ t=0.2 $.

\subsection{A comparative study of Einstein, Debye and $q$-deformed Einstein's Theory} 
 \begin{itemize}
 \item {\bf Einstein model:} In this model the nearest neighbour interactions between atoms 
 are neglected and each atom vibrates with the common frequency $\nu_E$ i.e. the phonon (quanta of collective 
 vibration of atoms inside the solid) frequency is assumed to be a constant, that is 
 independent of the wave vector $k(=\frac{2 \pi}{\lambda})$. 
 The specific heat as predicted by the Einstein model matches quite well when the temperature 
 of the solid is high (i.e. $T \gg \theta_E$) enough to disregard the nearest neighbour 
 atom$-$atom interaction. It predicts the Dulong$-$Petit value $3 R$ in the limit of very large 
 $T$ (i.e $T \to \infty$).  However, the model cannot 
 explain the sufficiently low temperature behaviour of the specific heat $C_V \propto T^3$, as the model 
 does not take care the nearest neighbour atom$-$atom interactions which need to be considered
 when the temperature $T$ of the solid is low. However, the model predicts $C_V \to 0$ 
 as $T \ll \theta_E$.
 
 \item {\bf Debye model:} In this model the nearest neighbour interactions between atoms are taken into consideration
 and each atom vibrates with a different frequency. Phonon (the quanta of the collective 
 vibration of atoms) frequency $\omega$ is assumed to vary linearly with the 
 wave vector $k(=\frac{2 \pi}{\lambda})$($\lambda$ denotes the wavelength), i.e, 
 $\omega=c_s \frac{2 \pi}{\lambda}$ where $c_s$ is the speed of sound. The so-called 
 Debye frequency $\omega_D$ is the frequency corresponding to the maximum allowed value 
 of $\frac{2 \pi}{\lambda}$ which is determined by the density of the material.
The Debye spectrum is an idealization of the actual situation obtaining in a solid. 
For low-frequency modes (popularly known as the acoustic modes) the Debye approximation works
quite well.  Some reasonable discrepancies have been there in the case of high-frequency modes 
(named as the optical modes) \cite{Patharia}. At sufficiently low temperature the Debye $T^3$ 
approximation works reasonably well. This leads us to the  fact that the Debye approximation is perfect when only long wavelength 
 acoustic modes are thermally excited. 
 On the other hand the energy of the short wavelength modes is too high for them to be 
 populated significantly at low temperatures. As a result,  $T^3$ approximation fails for them \cite{Kittel}.

 Debye's model tells us that due to interactions between each other the Einstein's oscillators cannot retain their fundamental frequency at a given 
 temperature; rather there will be a distribution of the frequencies of those oscillators with a maximum 
 allowed value of the frequency. 
 
  \item {\bf $q$-deformed Einstein model:} Unlike the Debye's model, where due to nearest neighbour 
  interaction, the atomic oscillators vibrates with a wide range of frequencies (ranging from zero to the 
  cut-off frequency $\nu_D$), in the $q$-deformed Einstein model, we introduce the nearest neighbour 
  interactions in the following way(see below), however maintain the spirit of the original Einstein model. 
  The interactions generate temperature fluctuations, which makes each of 3N 
  Einstein oscillator $q$-deformed and is vibrating with the same angular frequency
  $\nu_E$ as in the original Einstein (undeformed) model. The specific heat $C_V$  dependence on the temperature $T$ is found to be 
  in good agreement with the experimental data and also with the Debye's result. Thus we find that even 
  without considering more than one mode of vibration is being triggered, the model predicts 
  the $C_V$ vs $T$ graph quite well and is as good as obtained by considering the different
 modes of vibration as suggested by Debye's modification to the Einstein's theory. 
 
In Debye's model we have to assume the fact that the atomic oscillators vibrates with a 
  wide range of frequencies (ranging from zero to the 
  cut-off frequency $\nu_D$). Since the atom$-$atom interaction dominates over the 
  thermal agitation at low temperature, that excites more vibration modes. 
  Consequently in undeformed Debye model the cut-off frequency $\nu_D$ does not 
  remain constant anymore, the value $\nu_D$ increases at very low temperature 
  near $0 ~\rm{K}$ \cite{stewart, powell, tari}, which in nonextensive Tsallis scenario ($q$-deformed Einstein's scenario) not necessarily be the case, 
  i.e. $\nu_E$ can be temperature independent.
Considering the effect non-equilibrium conditions which arise due to nearest neighbour atomic interaction together with temperature 
fluctuation at very low temperature, certain level of impurities in the sample 
etc., along with the unique vibration mode as suggested by Einstein, the 
theoretical prediction is capable to explain the experimental data points. 

\end{itemize}

 {\textbf{Approximated compact form:}} Eq.(\ref{cvfinal}) can be converted to a more compact form using small deformation 
 approximation(also for the time being for this purpose $q$ is to be considered as a constant approximately, i.e., not a function of temperature and $\beta' \approx \beta$, see \ref{small_q}), i.e., small $\mid1-q\mid$ which states that, $e_q^a \cdot e_q^b \sim e_q^{a+b} $ (Eq.(\ref{smlmul})) and 
 $\left(e_q^a \right)^b \sim e_q^{ab} $ (Eq.(\ref{smlpwr})). Thus we find(see \ref{small_q}, Eq.(\ref{cvcompactapp}))
\bea
C_V =3 N k_B \frac{x^2 q \left(e_q^{-x}\right)^{2 q-1}}{\left[1-\left(e_q^{-x}\right)^q\right]^2}
\label{eq:cvcompact}
\eea

 For $q=1$(i.e., undeformed scenario) Eq.(\ref{eq:cvcompact}) replicates Eq.(\ref{eq:cveinstein}). Though Eq.(\ref{eq:cvcompact}) is an approximated compact analytical form  of Eq.(\ref{cvfinal}), it is not completely correct, as in our analysis we used $q$ as a function of temperature and also $\beta' \approx \beta$ is not very good approximation for a wide temperature range, specially at low temperatures. 
 
 Rather we can fit the data obtained using Eq.(\ref{cvfinal}) and get the specific heat of solid as a function of temperature, which is as follows
 
 \bea
 \frac{C_V}{3R}= A_1+\frac{A_2}{1+\left(\frac{x_0}{x}\right)^p}
 \label{eq:cvcompactexptfit}
 \eea
 with $x=\frac{\theta_E}{T}=\frac{h \nu_E}{k_B T}$, and the parameters $A_1=0.95$, $A_2=-0.96$, $x_0=4.12$ and $p=2.64$. Thus Eq.(\ref{eq:cvcompactexptfit}) can be approximated as
 
 \bea
 \frac{C_V}{3R} &\approx & 1-\frac{1}{1+ \left(4 \frac{T}{\theta_E}\right)^{2.7}} 
 \eea
 And at very low temperature, neglecting the higher order terms in temperature we obtain
 \bea
 \frac{C_V}{3R} \sim 1-\left\lbrace 1 - \left(4 \frac{T}{\theta_E}\right)^{2.7} \right\rbrace =\left(4 \frac{T}{\theta_E}\right)^{2.7}
  \label{eq:cvcompactexptfitlowtemp}
 \eea
 Interestingly from Eq.(\ref{eq:cvcompactexptfitlowtemp}) we find that the specific heat capacity at low temperatures predicted by $q$-deformed Einstein model(nonextensive Tsallis scenario), varies nearly as $T^3$($\sim T^{2.7}$) which is in nice agreement with experimental observation as well as with Debye's prediction.
\section{Conclusion}
We analyze the Einstein's theory for specific heat of solid in the generalized 
nonextensive scenario(Tsallis statistics).  We study the temperature fluctuation effect via the 
fluctuation in the deformation parameter $q$. At low temperature the Einstein's 
curve of the specific heat in the  nonextensive Tsallis scenario exactly lies on the 
experimental data points and matches with the Debye's curve. Considering only the temperature 
fluctuation effect(even without considering more than one mode of vibration is 
being triggered) we can achieve the desired result (i.e. $C_V$ vs $T$ curve) which 
is as good as obtained by considering the different modes of vibration of the 
solid as suggested by Debye. Finally, we find, by generalizing the Einstein's theory in nonextensive scenario(Tsallis statistics), that, a unique value of the Einstein temperature $\theta_E$, along with a temperature dependent deformation parameter $q(T)$, can well describe the 
 phenomena of specific heat of solid i.e. the theory is equivalent to Debye's theory with 
 a temperature dependent $\theta_D$.
\section*{ACKNOWLEDGEMENTS} 
\noindent The authors would like to thank Selvaganapathy J. for useful discussions and valuable suggestions. The work of PKD is supported by the SERB Grant No. EMR/2016/002651. One of the authors, Atanu, wants to thank Tuhin Malik and Debashree Sen for advice regarding tools.
\appendix

\section{Indicial properties of $q$-deformed exponential function for small deformation}

 From Eq.(\ref{eqn:expq}), keeping only first order in $(1-q)$,
\bea
e_q^a \cdot e_q^b &=& \left[1+(1-q) a \right]^\frac{1}{1-q} \cdot \left[1+(1-q) b \right]^\frac{1}{1-q} \nonumber \\
&=& \left[1+ (1-q) (a+b) + (1-q)^2 ab \right]^\frac{1}{1-q} \nonumber \\
& \simeq & e_q^{a+b} 
\label{smlmul}
\eea

 Similarly, neglecting higher order terms we get,
\bea
\left(e_q^a \right)^b &=& \left[1+(1-q) a \right]^\frac{b}{1-q} \nonumber \\
&=& \left[1+ (1-q) ab + \frac{b(b-1)}{2!}(1-q)^2 a^2+ \cdots \right]^\frac{1}{1-q} \nonumber \\
& \approx & e_q^{ab} 
\label{smlpwr}
\eea

For constant $q$ and $\beta' \approx \beta$, using Eqs.(\ref{smlmul}) and (\ref{smlpwr}) the following approximation can be obtained for Eq.(\ref{cvfinal})
\bea
 \frac{\sum_n \left(n+\frac{1}{2}\right)x \left\lbrace e_q^{-\left(n+\frac{1}{2}\right)x}\right\rbrace^q}{\sum_n \left\lbrace e_q^{-\left(n+\frac{1}{2}\right)x}\right\rbrace^q} &\approx &  \frac{\sum_n \left(n+\frac{1}{2}\right)x \left\lbrace e_q^{-nx}\right\rbrace^q \left(e_q^{-\frac{1}{2}x}\right)^q}{\sum_n \left\lbrace e_q^{-nx}\right\rbrace^q \left(e_q^{-\frac{1}{2}x}\right)^q} \nonumber \\
&\approx &  \frac{\sum_n \left(n+\frac{1}{2}\right)x \left\lbrace \left(e_q^{-x}\right)^q\right\rbrace^n }{\sum_n \left\lbrace \left(e_q^{-x}\right)^q\right\rbrace^n} \nonumber \\
&=& \frac{x}{2} \left[ \frac{1+\left(e_q^{-x}\right)^q}{1-\left(e_q^{-x}\right)^q} \right]
\eea
 
 Therefore Eq.(\ref{cvfinal}) takes the form(for constant $q$, and $\beta' \approx \beta$) as follows
\bea
\frac{C_V}{3 R} &=& \frac{x}{2} \left[ \frac{1+\left(e_q^{-x}\right)^q}{1-\left(e_q^{-x}\right)^q} \right] - x \frac{\partial}{\partial x} \left[ \frac{x}{2} \left\lbrace \frac{1+\left(e_q^{-x}\right)^q}{1-\left(e_q^{-x}\right)^q} \right\rbrace \right] \nonumber \\
&=& \frac{x^2 q \left(e_q^{-x}\right)^{2 q-1}}{\left[1-\left(e_q^{-x}\right)^q\right]^2}
\label{cvcompactapp}
\eea 

\protect\label{small_q}
\section{Constraints and Entropy Optimization in Tsallis Statistics}

 To impose the mean value of a variable in addition to satisfy the following fact
 \bea
 \int_0^{\infty} dx~p(x)=1
 \eea
 
$q$-deformed mean value of a variable $x$ is to be defined as \cite{Tsallis, Tsallis12}
\bea
<x>_q=\int_0^{\infty} dx~xP(x)=X_q
\eea

whereas, $P(x)$ is the Escort distribution and is defined as
\bea
P(x)=\frac{[p(x)]^q}{\int_0^{\infty} dx'~[p(x')]^q}
\eea

We immediately verify that $P(x)$ is normalized as well
\bea
\int_0^{\infty} dx~P(x)=\frac{\int_0^{\infty} dx~[p(x)]^q}{\int_0^{\infty} dx'~[p(x')]^q}=1
\eea

We can use these facts to optimize the generalized entropy $s_q$. In order to use the Lagrange's undetermined multiplier method to find the optimized distribution we define the following quantity
\bea
\Phi[p]=\frac{1-\int_0^{\infty} dx~[p(x)]^q}{q-1}- \alpha_q \int_0^{\infty} dx~p(x) -\beta_q \frac{\int_0^{\infty} dx~x[p(x)]^q}{\int_0^{\infty} dx~[p(x)]^q}
\eea
with $\alpha_q$ and $\beta_q$ as the Lagrange parameters. Therefore imposing the optimization conditions
\bea
\frac{\partial \Phi(p)}{\partial p}=0
\eea
Simplifying further we get 
\bea
p(x)=\frac{e_q^{-\beta_q \left(x-X_q \right)}}{\int_0^{\infty} dx' ~e_q^{-\beta_q \left(x'-X_q \right)}}
\eea

Now from the following two constraints
\begin{itemize}
\item $ \sum_i p_i=1$ (Norm constraint)
\item $ <\epsilon>_q = \sum_i \epsilon_i P_i= u_q$ (Energy constraint)
 
 with $P_i=\frac{p_i^q}{\sum_j p_j^q}$
\end{itemize}
 we obtain the distribution as follows
\bea
p_i=\frac{e_q^{- \beta_q \left( \epsilon_i - u_q \right)}}{\bar{z}_q}
\label{distribution_q}
\eea 
 
 with $ \bar{z}_q=\sum_i e_q^{- \beta_q \left( \epsilon_i - u_q \right)}$ and $\beta_q=\frac{\beta}{\sum_j p_j^q}$.

 Now 
 \bea
 s_q &=& k_B \frac{\sum_j p_j^q-1}{1-q} \nonumber \\
  \implies  \sum_j p_j^q &=& 1+ (1-q) \frac{s_q}{k_B}
 \eea

 Also
\bea
s_q = -k_B \ln_q \bar{z}_q= \frac{k_B}{1-q} \left( \bar{z}_q^{1-q} -1 \right)
\eea
 
\bea
\therefore \sum_j p_j^q = 1+ (1-q) \frac{\frac{k_B}{1-q} \left( \bar{z}_q^{1-q} -1 \right)}{k_B} = \bar{z}_q^{1-q} 
\eea
 
So now
\bea
\beta_q=\frac{\beta}{\sum_j p_j^q}= \beta \bar{z}_q^{q-1}
\eea 

 More useful and the convenient form of Eq.(\ref{distribution_q}) for application purpose, is given by \cite{Tsallis, Tsallis12}
 
\bea
p_i= \frac{e_q^{- \beta' \epsilon_i}}{Z_q}
\eea 

 with $ Z_q=\sum_i e_q^{- \beta' \epsilon_i}$ and $\beta'=\frac{\beta_q}{1+(1-q) \beta_q u_q}=\frac{\beta}{Z_q^{1-q}+(1-q) \beta u_q}$.
 
\protect\label{entropy_optimization}
\section{The temperature dependence of the deformation parameter $q$ in $q$-deformed Einstein's theory)}

\begin{figure}[!ht]
  \centering
  \subfloat[$q=1.03,1.05$]{\includegraphics[width=0.49\textwidth]{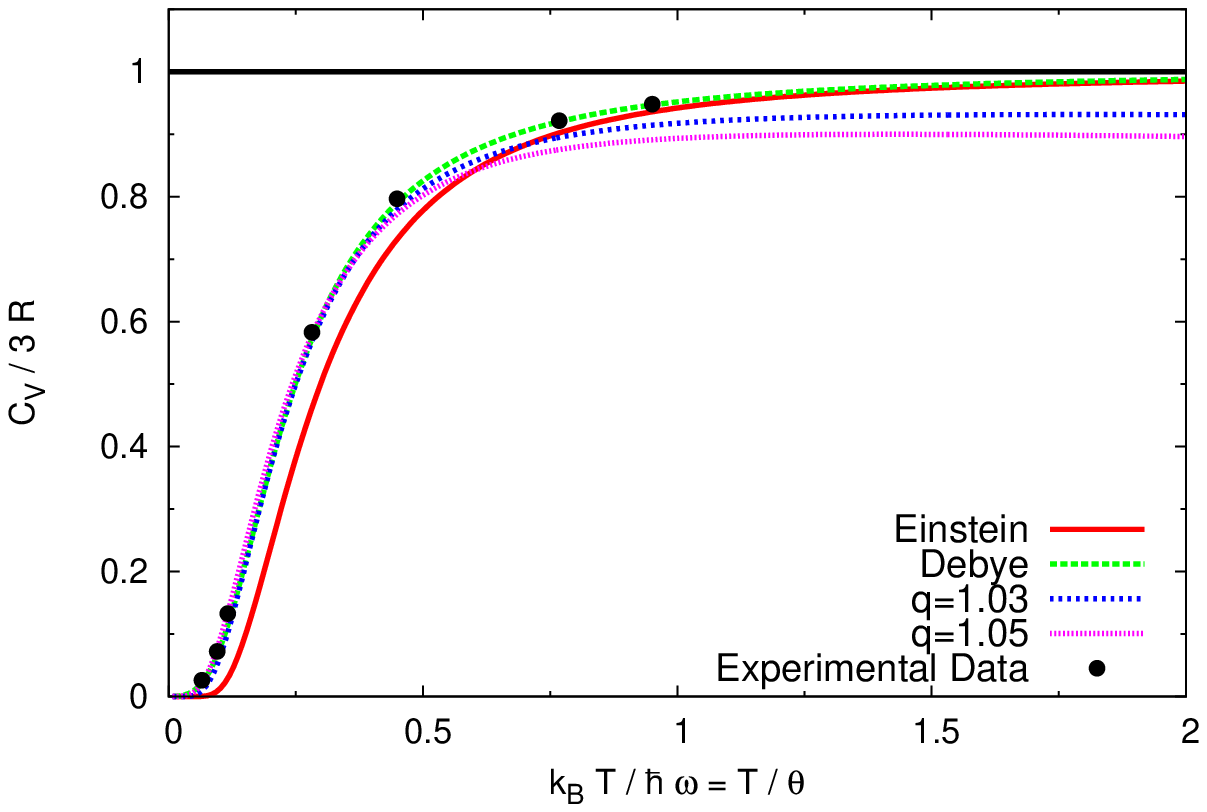}\label{fig:cve21}}
  \hfill
  \subfloat[$q=1.04,1.06$]{\includegraphics[width=0.49\textwidth]{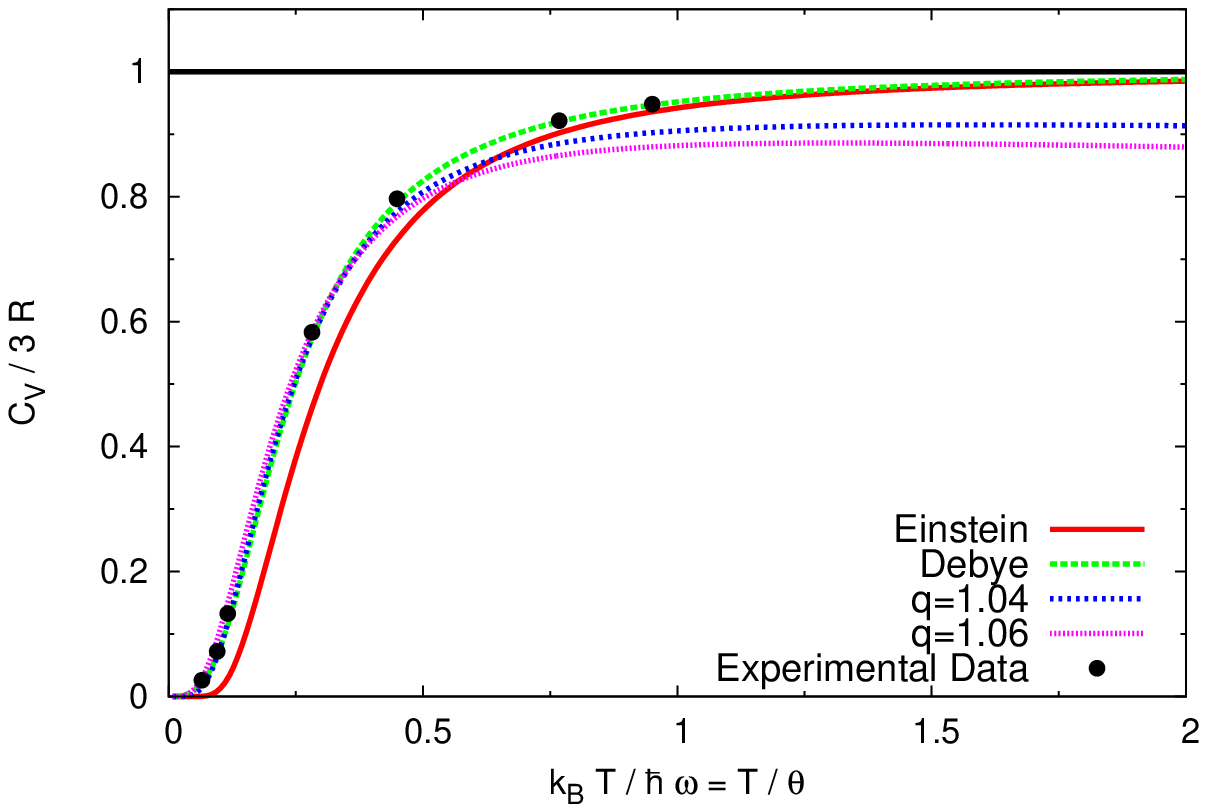}\label{fig:cve22}}
  \caption{\it Variation of $C_V/3R$ with temperature in the formalism of 
  Einstein's model, $q$-deformed Einstein modification is shown along with the 
  experimental data for copper \cite{white,bloom, stewart, Patharia}.}
\label{fig:cve2}
\end{figure}

In Fig.[\ref{fig:cve2}], we have plotted $\frac{C_V}{3R}$ as a function of 
$\frac{1}{x}(=\frac{T}{\theta_E})$ using Einstein's theory, Debye's modification and 
$q$-deformed Einstein's theory. Here we have used some constant value of the deformation 
parameter $q$ in Eq.(\ref{cvfinal}). In Fig.[\ref{fig:cve21}] we have showed the curves for $q=1.03,1.05$, while in Fig.[\ref{fig:cve22}] we set $q=1.04,1.06$.
 
 A close observation to the plots(Fig.[\ref{fig:cve2}]) establishes a temperature 
 dependence of the deformation parameter $q$. We found that, for comparatively high temperature i.e., 
 $ T \to 0.5\theta~ - ~ \theta$ the curve predicted by Debye's theory matches with the curve 
 obtained from small deformation(i.e., $q \sim 1.01 \rm{-} 1.03 $) where the thermal agitation 
 starts dominating over the atom$-$atom interaction. On the other hand at very small 
 temperature i.e., $T \sim 0.1~\theta \rm{-} 0.5~\theta $ the curve obtained from 
 comparatively large deformation(i.e., $q \sim 1.05 \rm{-} 1.07 $) lies on the curve 
 predicted by Debye's theory. For $ T > \theta$, $q \to 1.0 $(undeformed scenario) is enough explain the phenomena.
By fitting the experimental data points for Copper(Cu) with the curves obtained from different constant 
value of the deformation parameter $q$, we obtain Eq.(\ref{fit_q}).
\bea
q(T)=q_0 + A \exp {(- \frac{1}{t}\frac{k_B T}{\hbar \omega})}
\label{qtapp}
\eea 
 with $q_0=1.0, A=0.08 ~\rm{and}~ t=0.2 $. 

\protect\label{fit_Einstein}
%

%


\end{document}